\documentclass[fleqn,twoside]{article}
\usepackage{espcrc2}
\usepackage{graphicx}
\title{
{\vspace{-1.2em} \parbox{\hsize}{\hbox to \hsize
{\hss  \normalsize TRINLAT-03/03 , UPRF-2003-17}}} \\
Non-perturbative study of QCD on a $2+2$ anisotropic lattice}
\author{Giuseppe~Burgio\address[TCD]{The TrinLat Collaboration \\ 
        School of Mathematics, Trinity College, Dublin 2, Ireland}
         \thanks{Talk given by Giuseppe~Burgio},
        Alessandra~Feo\addressmark[TCD]\address[parma]
	{Dipartimento di Fisica, Universit\`a di Parma and 
	INFN Gruppo Collegato di Parma, Parco Area delle Scienze, 7/A, 43100 Parma, Italy},
        Mike~Peardon\addressmark[TCD]
        and
        Sin\'ead~M.~Ryan\addressmark[TCD]}       
\begin{document}
\begin{abstract}
We present preliminary results for the non-perturbative determination of
the parameters in a 2+2 anisotropic action. The parameters are required to restore Lorentz 
invariance in long-distance correlators, using the static inter-quark 
potential. Comparison with analytical results is made and further 
applications are discussed.
\end{abstract}

\maketitle

\section{INTRODUCTION}
QCD on 3+1 anisotropic lattices~\cite{Alford:1997nx,Morningstar:1997ff,Morningstar:1999rf} 
has proved to be a successful tool for the non-perturbative 
investigation of interesting physical phenomema including the glueball spectrum, $b\bar{b}$ states, 
gluon strings, the spectral density. 
The method is successful for processes where the hadronic final state is static in the center 
of mass frame but it is perhaps less advantageous for processes 
in which the hadron momentum in the final state is non-negligible compared to its mass. 
In such cases, independent fine spatial directions for each momentum
degree of freedom are needed to describe the system precisely. This increases 
dramatically the computational cost and, when considering an anisotropic approach,
the difficulties linked with tuning the parameters.

For processes where the final state hadron is strongly peaked at a momentum approximately 
equal to its mass, a 2+2 lattice discretization may provide a solution. 
It offers the possibility to extend the range of final state momenta accessible to 
lattice calculations and enhance the overlap with experimental data. 
In addition, the study of high-momentum glueball form factors could benefit from such a program. 

A 2+2 discretisation of the gauge action was proposed in Refs~\cite{Burgio:2002uq,Burgio:2003in} and 
the parameters in the action were determined at the one-loop level in perturbation theory. 
In this talk a non-perturbative determination of these parameters is discussed. A 2+2 
Wilson-like fermion action is discussed in Ref~\cite{feo}.
\section{YANG-MILLS THEORY ON A 2+2 LATTICE}
We start from the simple $1\times 1$ plaquette action
\begin{equation}
	S_{\rm g} = \beta\hspace{-0.1 cm}\sum_{x,\mu,\nu}c_{\mu\nu}[1 -\frac{1}{2 N_c}
			{\rm Tr} \, (P_{\mu\nu}(x) + P^\dagger_{\mu\nu}(x) ) ],
	\label{eqn:Sgluon}
\end{equation}
defined in a fixed hypercubic physical volume $V=L^4$. The ultraviolet cutoff is larger
 in the $z, \,t$ directions and $\xi = a_c/a_f$ is the asymmetry ratio. In the following, the coarse 
and fine lattice spacings are denoted $a_c$ and $a_f$ respectively. 
The coefficients $c_{\mu\nu}$ in Eq.~\ref{eqn:Sgluon} must be fixed by tuning their 
values so as to recover the correct continuum theory. This can be done either 
perturbatively~\cite{Burgio:2003in} or non-perturbatively~\cite{Alford:2000an}. 
The perturbative expansion is written
\begin{equation}
  c_{\mu\nu}= \left\{ \begin{array}{lll}
              \xi^2(1+\eta_{ff}^{(1)} g^2 +O(g^4))          & \mbox{f-f};\\
              1+\eta_{cf}^{(1)} g^2 +O(g^4)                 & \mbox{c-f};\\
              \frac{1}{\xi^2}(1+\eta_{cc}^{(1)} g^2 +O(g^4))& \mbox{c-c;}\end{array} \right.
\end{equation} 
and the calculation of the one-loop coefficients, $\eta_{\mu\nu}^{(1)}$ was the subject of 
Ref~\cite{Burgio:2003in}. We note in passing that the tree-level values of the $c_{\mu\nu}$ are
\begin{equation}
	c_{cc}=1/\xi^2, \,\, c_{cf}=1, \, \mbox{ and } \,c_{ff}=\xi^2.
	\label{eqn:cmunu-treelevel}
\end{equation}
A non-perturbative determination of these coefficients is the focus of this talk. 

\begin{figure}[htb]
	\includegraphics[scale=0.4]{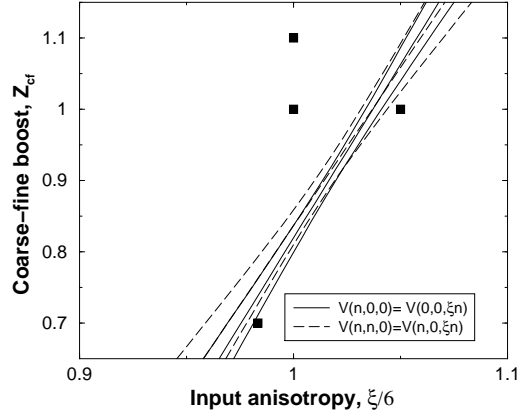}
	\caption{Degenerate constraint curves from the $SU(3)$ spatial potential.}
	\label{fig1}
\end{figure}

We choose to fix the coefficients to restore rotational invariance for the static inter-quark 
potential. The procedure is, in principle, the same as for a 3+1 discretisation. However,
for 3+1 any choice of the parameters leads to a well-defined, albeit unknown, scale ratio $\xi$.
In this case, a non-perturbative determination of $c_{\mu\nu}$ may be unnecessary if, as is 
true for some applications, one is not interested in the physical mass-scale ratio between the 
coarse and fine directions. 
If such a determination is necessary, both the coarse-fine and coarse-coarse Wilson loops 
must be measured.
The same is true for the 2+2 case, where the potential between spatially separated color sources, 
measured using a fine temporal axis determines just one of
the two free parameters in the action. The only drawback for the 2+2 discretisation compared to 
the 3+1 is that such a determination must be done {\it a priori}. 
Rotational invariance of the static inter-quark potential must be imposed for all 
combinations of source separation and transfer matrix axes. Alternatively, the slope of 
dispersion relations of physical (color singlet) states can be measured.
This requirement can be observed in perturbation theory by examining the static potential, 
given by
\begin{equation}
	V(R) = \lim_{T\rightarrow\infty} -\frac{1}{T}\log <W(R,T)> .
	\label{eqn:potl}
\end{equation}
Expanding Eq.~\ref{eqn:potl} at order $g^4$ and handling the resulting Bessel
functions for large $T$, it is easy to check that only the coefficients 
$\eta_{\mu\nu}^{(1)}$ carrying one time index survive. A fine temporal axis 
will therefore give no information about $\eta_{cc}^{(1)}$, since constraining potential ratios in
different directions leads to a constraint only on $c_{cf}/c_{ff}$ for 2+2 
and no constraint for 3+1. 
This constraint degeneracy is illustrated in Fig.~\ref{fig1}. The solid and dashed lines 
represent the constraints on the spatial potential: firstly that $V(n,0,0) = V(0,0,\xi n)$ 
and secondly that $V(n,n,0)=V(n,0,\xi n)$. The outer solid and dotted lines are the one sigma 
errors from a fit to these constraints. Within the errors
shown the lines are degenerate and cannot provide a non-perturbative determination of the 
$c_{\mu\nu}$.
Conversely, considering a coarse temporal axis results in no dependence on $\eta_{ff}^{(1)}$. For 
2+2, only the ratio, $c_{cc}/c_{cf}$ will be constrained. But now, this condition combined with the 
previous determination fixes the parameters uniquely.
For 3+1 this is the only meaningful condition. With a coarse temporal axis obtaining good fits 
can be more challenging. However, we find that with reasonable statistics good $\chi^2/d.o.f.$ is 
achieved. This non-perturbative determination of the $c_{\mu\nu}$ is work in progress.
\section{PRELIMINARY RESULTS}
\begin{figure}[htb]
	\includegraphics[scale=0.4]{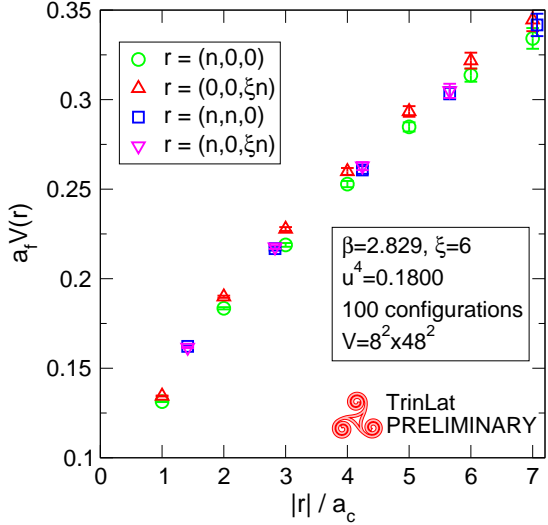}
	\caption{The $SU(3)$ static potential for $\xi = 6$ with tree-level coefficients 
	$c_{\mu\nu}$, given in Eq.~\ref{eqn:cmunu-treelevel}.}
	\label{fig2}
\end{figure}

As a first step we show that the tuning described in the previous section is actually not 
as onerous as it might sound. Fig.~\ref{fig2} shows the potential
in different directions, calculated using the tree-level coefficients. Even at the tree-level 
there is clearly no dramatic breaking of Lorentz invariance. This indicates that 
the tuning of parameters is mild and the scheme can be of practical use.

\begin{figure}[htb]
	\includegraphics[scale=0.4]{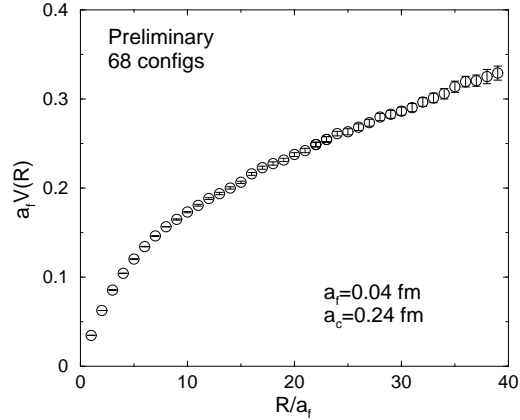}
	\caption{The $SU(3)$ static potential determined with the one-loop values of the 
	coefficients, $c_{\mu\nu}$~\cite{Burgio:2003in}. The potential is measured over 
	the range $R=0.04 - 1.6 fm$, for $\xi = 6$. 
	The errors on the points shown are smaller than the symbol size.}
	\label{fig3}
\end{figure}

To demonstrate how powerful this scheme can be, Fig.~\ref{fig3} shows the $SU(3)$ 
potential calculated for $\xi=6$ on a $8^2\times 48^2$ volume, 
in the range $R=0.04 - 1.6 fm$. These results were obtained after two days of workstation CPU 
time. 
The data show that the short distance term can already be calculated with very high precision. 
Longer runs and bigger volumes could easily allow one to reach even higher precision.
The method is therefore immediately applicable to a very high resolution determination of
the short-range inter-quark potential, which is of phenomenological interest.
\section{OUTLOOK AND DEVELOPMENTS}
We have shown how 2+2 lattice QCD can be relevant to useful physics.
Although parameter tuning is necessary, both the perturbative~\cite{Burgio:2003in} 
and the non-perturbative determinations are feasible. 
The results for the static inter-quark potential demonstrate that this scheme can 
be a useful tool for exploring energy ranges higher than those currently available to 
lattice simulations. 
In the gauge sector the scheme can naturally be applied to a calculation of glueball form factors, 
and this is under investigation. An extension to the quark action allows further interesting 
phenomenological applications and is described in Ref~\cite{feo}.

\section*{ACKNOWLEDGMENTS}
This work was funded by the Enterprise-Ireland grants SC/2001/306 and 307.

\end{document}